\documentclass[anonymous=true]{article}
\usepackage{spconf,amsmath,graphicx}
\makeatletter
\def\maketag@@@#1{\hbox{\m@th\normalfont\normalsize#1}}
\makeatother
\usepackage{cite}
\usepackage[shortcuts,acronym]{glossaries}
\usepackage{graphicx}
\usepackage{booktabs}
\usepackage{makecell}
\usepackage{multirow}
\usepackage{hyperref}
\usepackage{todonotes}
\usepackage{svg}
\usepackage{mathtools}
\usepackage{xcolor}
\everymath{\displaystyle}

\newacronym{cv}{CV}{Computer Vision}
\newacronym{bi}{BI}{Bayesian Inference}
\newacronym{dl}{DL}{Deep Learning}
\newacronym{dml}{DML}{Deep Metric Learning}
\newacronym{lar}{LAR}{Label-Aware Ranked}
\newacronym{ml}{ML}{Machine Learning}
\newacronym{rdis}{RDIs}{Range-Doppler Images}
\title{Label-Aware ranked Loss for robust People Counting using Automotive in-cabin Radar}
\address{Author Affiliation(s)}
\name{\begin{tabular}{c}Lorenzo Servadei$^{\star \circ}$ \qquad Huawei Sun$^{\star \circ}$ \qquad Julius Ott$^{\star \circ}$ \qquad Michael Stephan$^{\star \#}$ \qquad Souvik Hazra$^{\star}$ \\ 
\qquad Thomas Stadelmayer$^{\star \#}$ \qquad  Daniela Sanchéz Lopera$^{\star \circ}$ \qquad Robert Wille$^{\dagger}$ \qquad Avik Santra$^{\star}$\end{tabular}}

\address{$^{\star}$Infineon Technologies AG\\
$^{\circ}$ Technical University of Munich \\
$^{\dagger}$Johannes Kepler University Linz\\
$^{\#}$Friedrich-Alexander-University Erlangen-Nuremberg 
}
\begin{document}
\setlength{\abovedisplayskip}{4pt}
\setlength{\belowdisplayskip}{4pt}
\maketitle
\begin{abstract}
In this paper, we introduce the Label-Aware Ranked loss, a novel metric loss function. Compared to the state-of-the-art Deep Metric Learning losses, this function takes advantage of the ranked ordering of the labels in regression problems. To this end, we first show that the loss minimises when datapoints of different labels are ranked and laid at uniform angles between each other in the embedding space. Then, to measure its performance, we apply the proposed loss on a regression task of people counting with a short-range radar in a challenging scenario, namely a vehicle cabin. The introduced approach improves the accuracy as well as the neighboring labels accuracy up to 83.0\% and 99.9\%: An increase of 6.7\%  and 2.1\%  on state-of-the-art methods, respectively.
\end{abstract}
\section{Introduction}
\label{sec:intro}

The task of people counting is defined as predicting the number of people present in a given scene.
Its direct application in real-life scenarios made it particularly suitable for occupancy estimation~\cite{shih2015occupancy}, surveillance~\cite{zhang2007fast}, traffic management~\cite{10.1371/journal.pone.0186098}, and several other fields~\cite{tian2019automated, 9417826}. 
To this end, several solutions have been developed, particularly in the area of \ac{cv}, as in \cite{electronics10111293, 10.1007/978-3-030-34110-7_61, GAO20191}.
These approaches have shown positive results even on large crowds. Nonetheless, privacy preservation, as well as weather conditions independence are still major concerns for their implementation in real-life scenarios.
Accordingly, radar sensors offer a valid alternative that overcome the limitations mentioned above.

However, even though radar-based solutions are resistant to such limitations, their drawbacks are caused by issues such as low-resolution data which decrease the target-identification rate, missed detection as a result of occlusion, and unstable radar signal strength due to the superposition of reflections coming from various parts of the body. These challenges make the use of short-range and low-cost radars ineffective when used in conventional approaches for people counting in dense scenarios.

To overcome this, seminal works have been utilizing \ac{dl} approaches for advancing limitations given by traditional signal processing methods, as \cite{Aydogdu, Choi}. Although the work done shows advantages in terms of people counting performance, it still presents important challenges. In fact, the methods utilised there do not take advantage of the labels ranking implicit in the people counting task.


Additionally, the experimental scenarios presented are less prone to reflections and false targets w.r.t. small, close environments as automotive cabins, train carriages, etc.  
Nevertheless, these use-cases are very common in areas such as transportation and have been of particular importance for Heating, Ventilation, and Air Conditioning (HVAC) automated regulation.

In this paper, we present a novel loss function, namely the \ac{lar} loss, which leads to an improved \ac{dl} solution for people counting, performed on a challenging scenario inside a vehicle.
The presented loss function takes advantage of recent advancements in supervised \ac{dml}, a set of \ac{ml} methods whose goal is to learn such an embedding space in which similar sample pairs stay close while dissimilar ones are far apart.
Work done in this area, so far, has been focusing mostly on multi-class classification tasks, enhancing the decision margin among embedded vectors of unrelated classes \cite{8100196, 8578650, 9179805, deng2019arcface}. 
Instead, this contribution approaches the shaping of the embedding space in a regression problem. To this end, distance information among labels is exploited to have an increasingly ranked embedding space.
Additionally, to improve the robustness of the proposed solution, we enhance the generalization phase performance by adding an exponential moving average filter. Taking into account previous predictions enhance the stability of the estimated people counting.

As a result, the approach presented shows an increased accuracy up to 83.0\% and a neighboring labels accuracy (i.e. \emph{accuracy +/-}1) up to 99.9\%, thus respectively +6.7\% and +2.1\% w.r.t. the state-of-the-art methods on a real-world, complex people counting dataset.


\section{Background and Related Work}\label{sec:background}
In this section, we first review the radar signal processing methods utilized in related problem settings, and afterwards we analyse recent advancements in \ac{dml} which increase the performance of ML models.

\subsection{Radar Signal Processing}
\begin{figure}
     \centering
     \includegraphics[width = 0.9\linewidth]{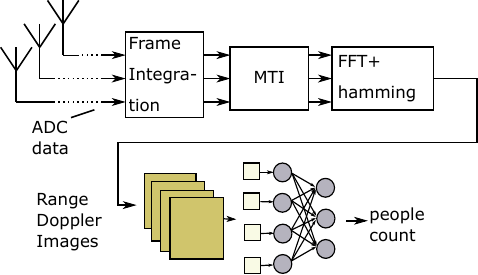}
     \caption{Block Diagram with Preprocessing}
     \label{fig:preprocessing}
     \vspace{-0.3cm}
 \end{figure}

Frequency Modulated Continuous Wave (FMCW) radars allow to estimate range, velocity, and angles of targets. For better interpretability, a preprocessing chain is typically used to extract some of these parameters from the sampled intermediate frequency signal before feeding it into the neural network. Different such preprocessing chains are given in \cite{santra2020deep}, in \cite{Zhang2020} they use \ac{rdis} as network inputs for object detection, while in \cite{Stephan2021} they work directly on the time domain data, but implicitly generate \ac{rdis} in the first network layer.
Fig.~\ref{fig:preprocessing} shows one such preprocessing chain.
First, a $N_S \times N_C$ 2D-dataframe is formed by acquiring the $N_S$ real samples from the intermediate frequency signal for each of the $N_C$ chirps, and stacking them column-wise. To increase the velocity resolution, a slow-time-frame with a larger observation period and therefore a higher velocity resolution is built, by integrating over the chirps of each frame, and by stacking $N_C$ of the resulting vectors to another 2D-dataframe of the same size. 
Afterwards, the mean values are subtracted along with the chirps as a moving target indication / high-pass-filter, to get rid of the Tx/Rx Leakage and to remove any completely static targets. As a last preprocessing step, both 2D-dataframes are transferred to frequency domain via 2D-FFTs. To reduce the sidelobe levels of each reflecting target, before doing the respective FFTs, the dataframes are multiplied with hamming windows of the respective sizes along the sample and chirp dimension. The 6-channel input to the neural network then consists of the real and the imaginary parts for the Range-Doppler Images of each antenna.
 
Although the pre-processing method showed good outcomes for different radar data use-cases, the final prediction outcome still suffers from instability. In fact, in tasks like people counting, where the count usually only changes by one or stays the same on a frame-by-frame basis, a temporal smoothing filter will help the network to provide a more stable and reliable prediction. An example of this is the Exponential Smoothing (ES), where the time data is smoothed with an exponential window function as seen in Eq.~\ref{eq:exponential}, where $x[k]$ is the current frame, $x_s[k-1]$ the smoothed frame from the last time-step $k$, and $x_s[k]$ the new smoothed frame.
\begin{equation}
\label{eq:exponential}
    x_s[k] = \alpha x[k] + (1-\alpha)x_s[k-1]
\end{equation}

While those methods have been important for many tasks related to radar data as people counting and tracking, the prediction mechanism has been relying more and more on \ac{dl}.

\subsection{Deep Metric Learning}
Embedding vectors are lower dimension vectors that represent high dimensional input data, such as images. \ac{dml} is an area of \ac{dl} that aims to learn data embedding vectors reducing the distance between samples of the same class, whereas the distance between samples of dissimilar classes is increased. In this paper, we consider the state-of-the-art \ac{dml} losses for benchmarking our proposed \ac{lar} loss.
Triplet loss~\cite{triplet} uses the Euclidean distance to measure the similarity between two embedding vectors. 
However, when the number of samples and classes becomes large, this loss becomes computationally expensive. 
To overcome this drawback, the Multiclass N-Pair (Mc-N-Pair) loss~\cite{npair} replaces the Euclidean distance with a dot product as a measurement to improve the computation efficiency. 
The Constellation loss~\cite{constellation} preserves the triplet structure, but it considers more negative classes than the Triplet loss during the update, for better convergence.
In the following, we provide an overview of these \ac{dml} losses and their properties.

\paragraph*{Triplet Loss}
Triplet loss~\cite{triplet} shown in Eq.~\ref{eq:2_triplet}, considers positive and negative pairs together. In every update step, there is an anchor $x_{i}^{a}$, a positive sample $x_{i}^{p}$, which has the same label as the anchor, as well as a negative sample $x_{i}^{n}$, which has a different label as the anchor. 
Then, the input triplet $\left\{ x_{i}^{a}, x_{i}^{p}, x_{i}^{n} \right\} $ is transformed to an embedding vector $\left\{ f_{i}^{a}, f_{i}^{p}, f_{i}^{n} \right\} $. 
In this case, the Euclidean norm between the transformed samples w.r.t the anchor sample is given by $E_p =\left\| f_{i}^{a}-f_{i}^{p}\right\|_{2}$ and 
$E_n =\left\| f_{i}^{a}-f_{i}^{n}\right\|_{2}$ for positive and negative samples respectively.
\begin{equation}
    \resizebox{.3\textwidth}{!}
    {
    $L_{tri}=\frac{1}{N}\sum_{i=1}^{N}{max \left( 0,E_p^{2}-E_n^{2}+m \right)}$,
    }
    \label{eq:2_triplet}
\end{equation}
\noindent where $m$ is the distance margin, and N is the batch size.

Triplet loss aims at minimising the distance between the anchor and the positive sample $E_p$, and maximizing the distance between the anchor and the negative sample $E_n$ at the same time. 

\paragraph*{Multiclass-N-pair Loss}
Triplet loss still suffers from slow convergence because in each update step only an anchor, a positive, and a negative sample are considered. Mc-N-Pair loss~\cite{npair} extends the Triplet loss, using all the labels for each update, as shown in Eq. \ref{eqn:mc-n-pair}. 
\begin{equation}\label{eqn:mc-n-pair}
    \resizebox{.4\textwidth}{!}
    {$
    L_{Mc-Npr}=\frac{1}{N}\sum_{i=1}^N{\begin{array}{c}
	log \left( 1+\sum_{j\ne i}^{}{\exp \left( f_{i}^{T}f_{j}^{n}\,\,-\,\,f_{i}^{T}f_{i}^{p} \right)} \right).
\end{array}} $}
\end{equation}
This property improves the convergence, but if the number of labels is too high, it leads to computational inefficiency. This can further affect the training of the network.

\paragraph*{Constellation Loss}
The Constellation loss~\cite{constellation}, shown in Eq. \ref{eqn:constellation_loss}, combines the advantages of Mc-N-Pair loss and Triplet loss. Compared to Triplet loss, it takes more negative labels into account, while it follows the formulation of the Mc-N-Pair loss.
\begin{equation}\label{eqn:constellation_loss}
    \resizebox{.4\textwidth}{!}
    {$
    L_{cons}=\frac{1}{N}\sum_{i=1}^N{\begin{array}{c}
	log \left( 1+\sum_j^{K}{\exp \left( f_{i}^{aT}f_{j}^{n}\,\,-\,\,f_{i}^{aT}f_{j}^{p} \right)} \right),\\
\end{array}}$}
\end{equation}
where $K$ is an optional hyperparameter that indicates how many negative labels are considered for each update. 

Although the presented DML losses have been used in multiple tasks, to the best of our knowledge no loss takes distances among labels into account for ordering the embedding space and improving the regression prediction. 


\section{Proposed Approach}

To address these issues, in this section we present the proposed loss function, namely the LAR loss, and successively demonstrate its theoretical foundations and properties. To this end, we show that it minimises when the rank among labels (i.e. number of people in a scene) is preserved, and the angles among embeddings of different labels are uniform.

LAR takes inspiration from the Constellation loss but additionally takes advantage of the labels' information to reproduce their ranking in the embedding space, thus enhancing the prediction capabilities of the models. 
The  LAR loss is presented in Eq. \ref{eqn:label_multi}.

\begin{equation}\label{eqn:label_multi}
    \resizebox{.44\textwidth}{!}
    {$
    L_{LAR}=\frac{1}{N}\sum_{i=1}^{N}{\begin{array}{c}
	\log \left( 1+\sum_{j \ne i}{\exp \left( \log(\Delta_l) f_{i}^{aT} f_{j}^{n}\,\,-\,\,f_{i}^{aT} f_{j}^{p} \right)} \right),
\end{array}}$}
\end{equation}
where
\begin{equation}\label{eqn:multipl}
    \resizebox{.25\textwidth}{!}
    {
    $\Delta_l=min\left(|l_a - l_n|, \left|L-|l_a-l_n|\right| \right)$
    }.
\end{equation}

The loss uses the multiplier $\log(\Delta_l)$ to regulate the ranking of the labels.
Here, $l_a$ is the label of the anchor,
$l_n$ is the label of the current negative sample
and $L$ is the number of different labels. Here, $f_i$ identifies the embedding of the input sample.
The multiplier assigns smaller values to neighbouring labels and establishes a distance metric among labels. The logarithm function is applied to it, as it is monotonically increasing and adds numerical stability.


In LAR, we use normalised feature vectors, i.e. $ \langle f_i,f_j \rangle = \cos(\theta)$, thus our loss operates on the angles between the feature vectors.  
Similar to Mc-N-Pair and Constellation loss, in the LAR loss, the embedding vectors of the same label are pushed to an angle of $\theta=0$, which minimises the loss.
In the following, we show the properties of our LAR loss.
Regarding different labels, we show that having ranked labels and uniform angles minimises the LAR loss.

\noindent By Jensen's inequality for convex functions, i.e.
\scriptsize
\begin{equation}
    \frac{\sum\limits_{i=1}\limits^I e^{\cos(\theta_i)}}{I} \geq e^{\frac{\sum\limits_{i=1}\limits^I \cos(\theta_i)}{I}},
\end{equation}
\normalsize
without loss of generality, we can examine only the angles among the labels that minimise the $\sum \cos(\theta_i)$.
Furthermore, the cosine is minimal at $\theta=\pi$. Thus, in the case of an even $L$, we expect the angle between labels with the highest multiplier to be $\pi$.  This can be only achieved by a uniform angle $2\pi/L$ among all the  labels. 
In the case of an odd $L$, we start with the assumption that our points lie on the unit hyper-sphere in uniform angles. Then, we show that the sum of cosines with uniform angles between classes is always smaller than the same sum where one of the points is shifted by an $\epsilon$ on the circle which fulfils $0<\epsilon<2\pi/L$. Shifting only one angle by $\epsilon$ is the smallest perturbation possible and by showing that this will increase the loss, we cover the cases of multiple perturbations as well. 
Since not all cosine terms are affected by the shift of one point, Eq. \ref{eqn:proof_odd1} states only the difference. 

\scriptsize
\begingroup
\setlength\abovedisplayskip{0pt}
\setlength\belowdisplayskip{0pt}
\begin{multline}\label{eqn:proof_odd1}
    \sum_{l=3}^{L} 2 \sum_{j=1}^{\frac{l-1}{2}} 2\log(j)\cos \left(j\frac{2\pi}{l} \right)  \\
    < \sum_{l=3}^{L} \sum_{j=1}^{\frac{l-1}{2}} 2\log(j)\cos \left(j\frac{2\pi}{l} - \epsilon \right) + 2\log(j)\cos \left(j\frac{2\pi}{l} + \epsilon \right),
\end{multline}
\endgroup
\normalsize
Now we can apply the properties of the cosine and obtain
\scriptsize
\begingroup
\setlength\abovedisplayskip{0pt}
\setlength\belowdisplayskip{2pt}
\begin{equation}\label{eqn:proof_odd2}
    \sum_{l=3}^{L} 2 \sum_{j=1}^{\frac{l-1}{2}} 2\log(j)\cos \left(j\frac{2\pi}{l} \right) 
    <\sum_{l=3}^{L} 2cos \left (\epsilon \right)\sum_{j=1}^{\frac{l-1}{2}} 2\log(j)\cos \left(j\frac{2\pi}{l} \right).
\end{equation}
\endgroup

\normalsize
By the symmetry of the cosine, the uniform angles and the ordering of our multiplier, {\small$\sum_{j=1}^{\frac{l-1}{2}} 2\log(j)\cos \left(j \frac{2\pi}{l}\right) < 0$} holds for any odd $l 
\geq 3$. Hence, we can divide by the inner sum which yields
\scriptsize
\setlength\abovedisplayskip{2pt}
\setlength\belowdisplayskip{2pt}
\begin{equation}\label{eqn:proof_odd3}
    \begin{array}{l}
    \sum\limits_{l=3}\limits^{L} 1 > \sum\limits_{l=3}\limits^{L} \cos(\epsilon).
  \end{array}
\end{equation}
\normalsize
As known, $\cos(0)=\cos(2\pi)=1$ are the maxima of the cosine, and these upper bounds are never reached by any $\cos(\epsilon)$ with $0<\epsilon<2\pi /L$ and $L \geq 3$. 
For an even $L$, the inequality changes to
\scriptsize
\setlength\abovedisplayskip{0pt}
\setlength\belowdisplayskip{2pt}
\begin{multline}
    \sum_{l=4}^{L} 2 \sum_{j=1}^{\frac{l}{2}-1} 2\log(j)\cos \left( j\frac{2\pi}{l} \right) + \frac{l}{2}\cos(\pi) \\
    < \sum_{l=4}^{L} 2\cos(\epsilon)\sum_{j=1}^{\frac{l}{2}-1} 2\log(j)\cos \left(j\frac{2\pi}{l} \right) +\frac{l}{2}\cos(\pi - \epsilon).
\end{multline}
\normalsize
Since, $\cos(\pi) < \cos(\pi - \epsilon)$ is always true for any $0<\epsilon<2\pi / L$ we can exclude it from the inequality which yields
\scriptsize
\setlength\abovedisplayskip{0pt}
\setlength\belowdisplayskip{2pt}
\begin{equation}
    \sum_{l=6}^{L} 2 \sum_{j=1}^{\frac{l}{2}-1} 2\log(j)\cos \left(j\frac{2\pi}{l} \right) 
    <  \sum_{l=6}^{L} 2\cos(\epsilon)\sum_{j=1}^{\frac{l}{2}-1} 2\log(j)\cos \left(j\frac{2\pi}{l} \right).
\end{equation}
\normalsize
Now the same reasoning from the odd case applies here. 
This shows that our loss is minimal for uniform angles and when the ranking is preserved.
\begin{figure}
     \centering
     \includegraphics[width = 0.7\linewidth]{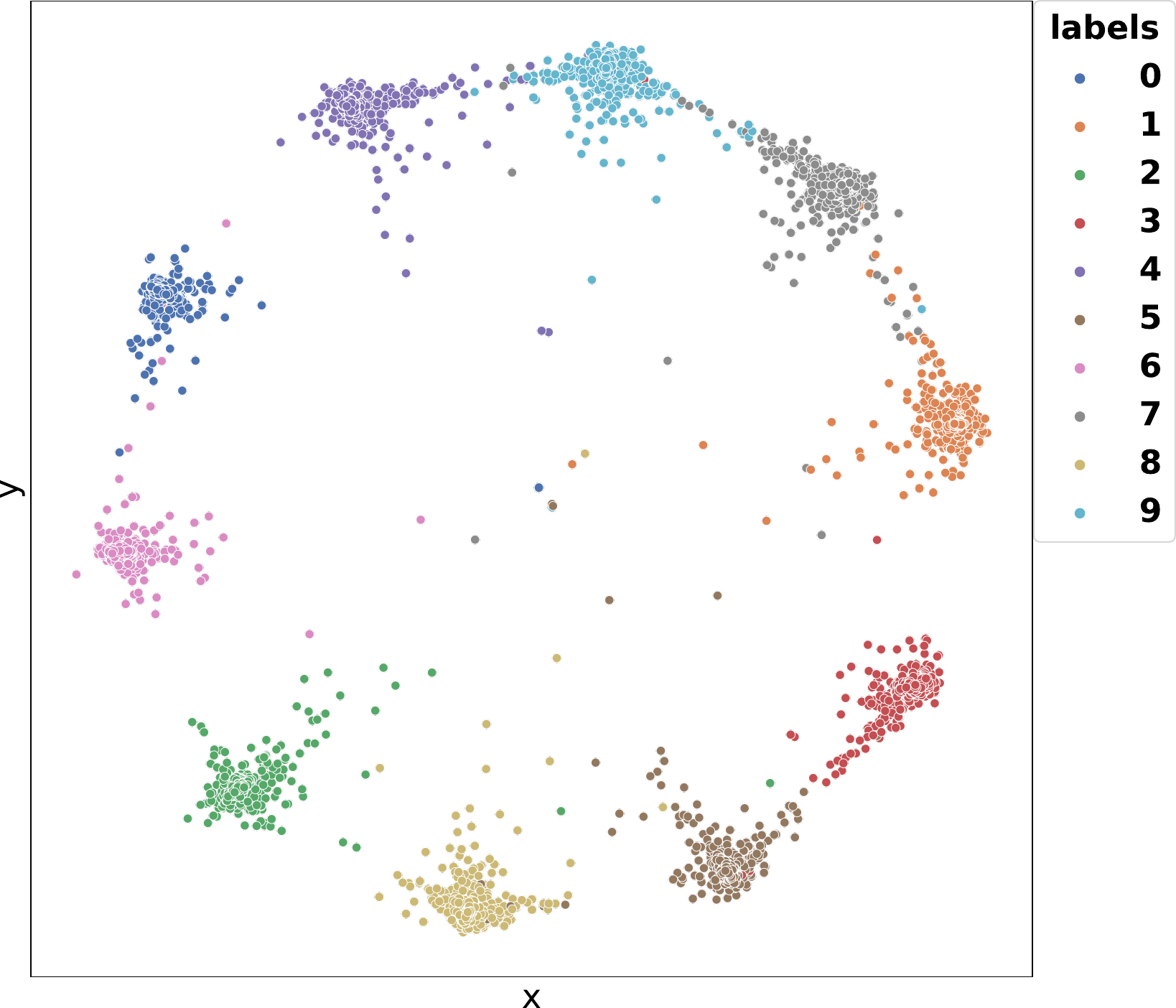}
     \caption{Constellation Loss Embeddings on MNIST}
     \label{Fig: Non_Ranked}
     \vspace{-0.3cm}
 \end{figure}
 
 \begin{figure}
     \centering
     \includegraphics[width = 0.7\linewidth]{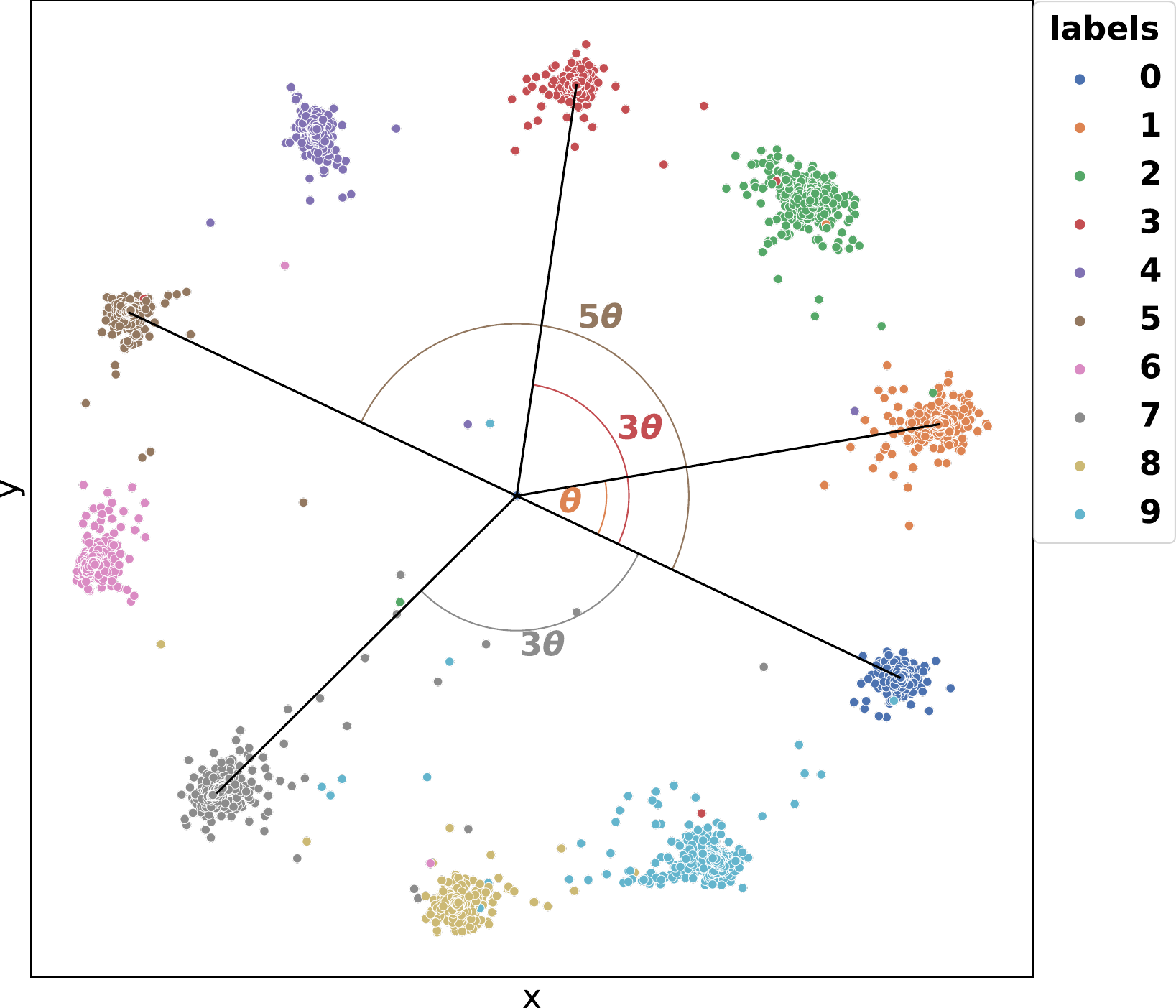}
     \caption{LAR Loss Embeddings on MNIST}
     \label{Fig: Ranked}
     \vspace{-0.3cm}
 \end{figure}

To show the effect of the LAR loss onto a simple dataset, we applied it to the MNIST dataset \cite{deng2012mnist}. The embedded space resulting from the Constellation loss is presented in Fig. \ref{Fig: Non_Ranked}, while the effect of the LAR loss is shown in Fig. \ref{Fig: Ranked}. There we observe that the LAR loss ensures uniform angles and ordering between the class mean and shows higher discriminative power between the classes.

To encourage stability on the people counting task, we add ES to stabilize the inference output of the network.

In this repository \footnote{LAR: \href{https://github.com/2Geeks2/LabelAwareRanked-Loss.git} {\textcolor{blue}{https://github.com/2Geeks2/LabelAwareRanked-Loss.git}}}, we show the effect of LAR on uniform angles and labels ranking, on a randomly generated dataset and MNIST.

In the next section, we apply the proposed LAR loss on a real-world dataset and show how the implicit ordering of the embedding would help in a regression problem. In fact, by constructing an ordered representation on the embedding space, we implicitly express the ranking of the labels and support the model in the final prediction.

\section{Experimental}
\label{sec:typestyle}

In this section, we first review the implementation settings, successively we benchmark our proposed loss function with and without temporal smoothing in an ablation study.

\subsection{Implementation Settings}
In the implementation, we used PyTorch v1.8.0.\textsuperscript{\texttrademark}- GPU v2.4.0 with CUDA\textsuperscript{\textregistered} Toolkit v11.1.0 and cuDNN v8.0.5. As a processing unit, we used the Nvidia\textsuperscript{\textregistered} Tesla\textsuperscript{\textregistered} P40 GPU, Intel\textsuperscript{\textregistered} Core i7-8700K CPU, and DIMM 16GB DDR4-3000 module of RAM. 
In order to count people, one Infineon's XENSIV\textsuperscript{\texttrademark} 60 GHz sensor has been utilized, on the internal-upper side of the front window of the vehicle cabin.
The radar input data has been pre-processed as explained in Section \ref{sec:background}, and results in 95000 frames of scenes of people counting inside a vehicle, performed on zero to five people, and divided into recordings with an average length of 350 frames. The frames are recorded with a frame rate of $10$ Hz.
The dataset has been split into training and test set, dividing it into 76000 (training) and 19000 (testing) frames. In order to benchmark state-of-the-art losses, 
we create a smart batch structure, as mentioned in \cite{npair}. This means, every batch contains two samples per label, thus accounting for a batch size of 12. Additionally, we take advantage of an equal number of samples for each label. This corresponds to around 16k samples for each class.
For the people counting task, we utilize a network with the same architecture as proposed in the encoder of \cite{Aydogdu}, composed of three convolutional layers with ReLU activation and a pooling layer after each convolutional layer. Each of the convolutional
layers has 32 feature maps and a kernel size of $3 \times 3$. As the last layer we use a fully connected ReLU layer where we round the output for the prediction.

\subsection{Ablation Study}
In order to show the outcome of our approach, we implement the proposed methods in an ablation study.
To this end, we analyse different losses and benchmark them on the aforementioned people counting dataset, as shown in Table \ref{Table:Benchmark}. Here, both \emph{accuracy} and \emph{accuracy +/-}1 are shown. While \emph{accuracy} points out the overall accuracy on the test set, the \emph{accuracy +/-}1 score takes into account the two neighboring labels (for the label 0 and 5, only one is considered).
To obtain the accuracy measure, we round the regression predictions obtained using the Mean Squared Error (MSE) loss.
In our benchmark, we first compare our LAR loss to the state-of-the-art \ac{dml} losses. Additionally, we compare the same losses by enhancing and stabilizing their outcome using ES.

\begin{table}[htb]
\centering
\caption{Benchmark of DML Losses}
\begin{tabular}{@{}ccc@{}}
\toprule
                             & \textbf{Accuracy}  & \textbf{Accuracy +/-1 }   \\ \midrule
\makecell{MSE}   & 68.6\%     & 94.5\%              \\ 
{MSE + Triplet } & 70.6\%   & 95.3\%   \\  
\makecell{MSE + Mc-N-Pair } & 68.9\%   & 95.7\%  \\   
\makecell{MSE + Constellation } & 74.6\%   & 97.2\%  \\ 
\makecell{\textbf{MSE + LAR (Ours)} } & \textbf{80.8\%}   & \textbf{98.5\%}  \\
\midrule  
\makecell{MSE + ES } & {71.9\%}   & {96.4\%}  \\
\makecell{MSE + Triplet + ES } & {72.7\%}   & {97.1\%}  \\
\makecell{MSE + Mc-N-Pair + ES } & {72.8\%}   & {97.3\%}  \\
\makecell{MSE + Constellation + ES } & {76.3\%}   & {97.8\%}  \\
\makecell{\textbf{MSE + LAR + ES (\textbf{Ours}) }} & \textbf{83.0\%}   & \textbf{99.9\%}   \\   \midrule
\bottomrule

\end{tabular}
\vspace{-4mm}
\label{Table:Benchmark}
\end{table}

Table \ref{Table:Benchmark} shows the results for our proposed approach against the state-of-the-art. Our proposed method shows an accuracy of 80.8\% and a +/-1 accuracy of 98.5\%, respectively +6.2\% and +1.3\% towards the second best performing loss (i.e. MSE + Constellation loss). Adding the ES, our method reaches an accuracy of 83.0\% and a +/-1 accuracy of 99.9\%, respectively +6.7\% and +2.1\% towards the second best performing loss (i.e. MSE + Constellation loss + ES).
By ranking the latent space and ordering the embedding, we show the advantages in prediction performance, even when the neighboring labels are considered.

\section{Conclusion}
In this work, we formulated a novel loss function, namely LAR loss. First demonstrated that it minimises when samples are ranked in label orders in the embedding space, and they are separated by uniform angles between labels. Afterward, we show how these properties can help in the people counting task on a difficult scenario of a vehicle cabin. To this end, we first train a CNN with LAR and MSE losses. Afterward, we increase the robustness of the prediction using ES. As a result, our approach presents an increment in accuracy and accuracy +/- 1 of respectively +6.7\% and +2.1\% towards the second-best performing loss.

\bibliographystyle{IEEEbib}
\bibliography{strings,refs}

\end{document}